# Survey on Neighbor Discovery and Beam Alignment in mmWave-Enabled UAV Swarm Networks

Muhammad Morshed Alam and Sangman Moh
Dept. of Computer Engineering
Chosun University
Gwangju, 61452 Korea
morshed@chosun.kr, smmoh@chosun.ac.kr

## ABSTRACT

Millimeter wave (mmWave)-enabled unmanned aerial vehicle (UAV) swarm networks (UAVSNs) can utilize a large spectrum of resources to provide low latency and high data transmission rate. Additionally, owing to the short wavelength, UAVs equipped with large antenna arrays can form secure narrow directive beam to establish communication with less interference. However, due to the high UAV mobility, limited beam coverage, beam misalignment, and high path loss, it is very challenging to adopt the mmWave communication in UAVSNs. In this article, we present a comprehensive survey on neighbor discovery and beam alignment techniques for directional communication in mmWave-enabled UAVSNs. The existing techniques are reviewed and compared with each other. We also discuss key open issues and challenges with potential research direction.

## 1 INTRODUCTION

The collaborative unmanned aerial vehicles (UAV) swarm networks (UAVSNs) have many applications in both military and civilian areas. In UAVSNs, a mission can be successfully achieved by efficient task allocation and intelligent formation control. The task allocation algorithm allocates the target task and role to each UAV [1]. The formation control algorithm maintains the swarm topology and perform obstacle avoidance via collaborative motion planning [2].

A UAVSN can be deployed to collect data from the Internet of things (IoT) devices and to perform real-time sensing by transmitting high-resolution video to the base station (BS). The UAVSN can also be deployed to function as an aerial BS to provide communication coverage to the ground users (GU) [3]. It can deliver high-capacity communication links to the GUs, IoT devices, and the remote BS by acting as a mobile relay back hauler during emergency [4]. To meet the low latency, high data rate, and secured interference-limited data transmission in above-mentioned UAVSN applications, the utilization of the plentiful spectrum resources provided by millimeter Wave (mmWave) can deliver a potential solution.

Owing to the advantage of the short wavelength of mmWave, it is possible to design a compact antenna array to attach with size-, payload-, and energy-constrained UAVs. It can form a narrow directive beam toward a specific target UAV to establish long-distance directional communication with desired signal-to-interference-plus-noise ratio (SINR). The concept of beamforming is implemented to radiate maximum energy toward a specific target direction. The beamforming produces one main lobe and some side lobes. The side lobes are unavoidably generated in an unwanted direction because it is not possible to entirely cancel the radiation in an unwanted direction. The width of the main lobe is known as beamwidth as illustrated in Fig. 1. The main objective of beamforming is to maximize the link SINR.

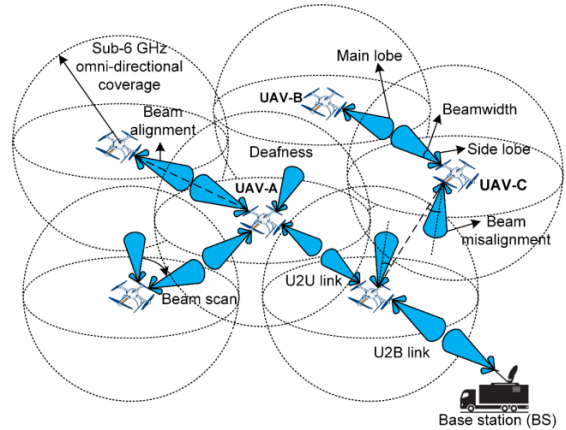

**Figure 1: An example of mmWave-enabled UAV swarm network.**

UAV to UAV (U2U) links are mostly treated as LoS links and produce free space paths due to high altitude. Similarly, UAV to GU (U2G) or UAV to BS (U2B) links are dominated by LoS as UAVs can adjust their three-dimensional (3D) position. However, considering the high mobility of UAVs in 3D space and wind disturbance, the narrow directive beam has the limited coverage to maintain the desired SINR. In addition, due to the limited coverage of narrow directive beams, the real-time neighbor discovery process becomes very challenging in UAVSNs. Although the mmWave quasi-omni-directional antenna beams can provide full space beam coverage, they have a very limited coverage radius as the antenna gain is reduced [5].

In the mmWave-based directional communication in UAVSNs, the link SINR depends not only on optimal allocation of resources but also on the beam alignment error and U2U distance. Due to mobility and hovering fluctuations, the beam misalignment can cause severe link outage [4], [6]. Simultaneously, real-time neighbor discovery is required to perform low-latency relay operation while incurring minimal overhead. To maintain the desired link SINR in UAVSNs, mobility control, beam alignment, and beamwidth control are necessary between UAVs by using an intelligent formation control and performing the beam steering.

The formed beamwidth can be adaptively controlled by activating and deactivating a particular sub-array of AE. This can be achieved by the required SINR, the changes in mobility, and the angle of arrival (AoA)/angle of departure (AoD) of signals from neighbor UAVs [6–8]. In [6] and [9], an analytical channel model was studied for mmWave-enabled rotary wing UAVs under the hovering fluctuations. Through the trade-off between antenna beamwidth and directivity gain, they found that the adaptive beamwidth control is necessary to minimize the link outage in hovering fluctuations. In a dynamic environment, the hovering fluctuations and the changes in relative mobility can create severe beam misalignment problem even though the high directivity gain ensures long-distance propagation as shown in Fig. 1. Thus, to ensure a reliable link, adaptive beamwidth control is required under a threshold U2U distance.

In the mmWave-enabled UAVSNs, the high UAV mobility, limited beam coverage, beam misalignment, and high path loss make the directional communication challenging. In this paper, we survey the neighbor discovery techniques (NDTs) and beam alignment techniques (BATs) for mmWave-enabled UAVSNs. The existing NDTs and BATs are comprehensively reviewed and then qualitatively compared with each other in terms of major features, advantages, and limitations. Critical open issues and research challenges are also addressed.

The rest of the paper is organized as follows: In Section 2, the basic concepts of mmWave communication in UAVSNs are summarized by discussing smart mmWave antennas, beamforming, topology formation, and challenges in directional communication. In Section 3, we present the different NDT and BAT techniques. In Section 4, important open issues and challenges along with potential research direction are discussed for the future mmWave-enabled autonomous UAVSNs. Finally, this paper is concluded in Section 5.

## 2   mmWAVE COMMUNICATION IN UAVSNS

In this section, the mmWave antennas, beamforming, and topology formation in mmWave-enabled UAVSNs are briefly overviewed along with their advantages and limitations.

### 2.1   Antennas in mmWave Communication

To compensate for the higher path loss of mmWave, the compact higher AE-based antenna array can be used for narrow directional beamforming. As shown in [6], the antenna directivity gain is increased with increasing the number of active AEs. Simultaneously, with increasing the antenna gain, the beamwidth is decreased. It is vice versa if the active AE is reduced by turning it off. The AE can be designed in different geometry shapes for beamforming such as uniform linear array (ULA), uniform planar array (UPA), and conformal array (CA). In ULA, the AE is placed linearly with uniform space on the z-axis [6]. In UPA, the AE can be placed uniformly in rectangular or square shapes. In CA, the AE can be designed as circular and cylindrical shapes [8]. The ULA has a simple linear structure and is less complex. However, the ULA can give beam alignment only in a single plane. ULA is mostly supported by the rotary wing UAVs as they can hover in a fixed position and maintain a fixed altitude.

Similarly, the UPA is relatively less complex and can provide two-dimensional (2D) beam coverage [9]. The UPA is suitable to attach to both rotary-wing and fixed-wing UAVs. Because fixed-wing UAVs have a large platform size, the UPA can be attached both up and downside of the wing for 3D full or half-space beam coverage [10]. The UAVs may require an extra payload to attach the ULA and UPA-based antenna array. In contrast, in CA (especially, the cylindrical-shaped CA) can provide 3D full space beam coverage. In CA, the AE can be conformed to a curved surface (i.e., aircraft body or wings). Thus, CA provides more degree of freedom for antenna geometry design and delivers better aerodynamic performance. The CA is very suitable for fixed-wing UAVs because the AE can be conformed in cylindrical shapes in the body of the aircraft such as the fuselage [8]. As a result, it does not require any extra payload for fixed-wing UAVs. Nevertheless, beamforming and beam steering are more complex in CA due to the irregular geometry design of AE.

### 2.2   Beamforming in mmWave Communication

The beamforming techniques are usually categorized as digital, analog, and hybrid. In digital beamforming, each AE is connected to a radio-frequency (RF) chain. Here, beamforming is performed in the baseband by using digital precoding. Digital beamforming supports multi-stream transmissions, provides higher spectral efficiency, and it can separate the simultaneously received signal from multiple directions. However, it consumes very high energy because it allocates a dedicated RF chain to each AE, which requires complex hardware structures such as analog to digital converter and digital to analog converter. In contrast, analog beamforming requires only a single RF chain, and it requires switches or phase shifters in the analog domain to perform the beam and beam steering. Thus, it has less hardware complexity and provides higher energy efficiency. However, it supports only single-stream transmission and offers low flexibility as only the phase of the signal generated by each AE can be controlled during beamforming.

As a result, hybrid beamforming is getting attractive to achieve the advantages of both analog and digital beamforming by addressing the trade-off between energy efficiency and spectral efficiency. Hybrid beamforming requires fewer RF chains compared to digital beamforming to reduce the energy consumption, hardware complexity, and deployment cost. Hybrid beamforming can be further classified as fully connected and partially connected beamforming. In fully connected hybrid beamforming, less RF chains are used compared to pure digital beamforming, and each RF chain is connected to all the AEs via phase shifters, resulting in the increased hardware complexity. However, it can deliver full beamforming gain compared to partially connected structure. In the partially connected hybrid beamforming, AEs are grouped into different sub-arrays. Each sub-array is connected to an RF chain through phase shifters. Thus, partially connected hybrid beamforming has less hardware complexity compared to fully connected architecture. Apart from the above beamforming techniques, recently intelligent reflecting surfaces (IRS) are used to form passive beamforming by controlling the phases and amplitudes of reflected signals. The IRS-based passive beamforming delivers higher energy efficiency and lower hardware cost.

### 2.3 Topology Formation in mmWave-Enabled UAVSNs

According to the role of UAVs, we classify the UAVSN topology as flat mesh topology [10] and leader-follower topology [7], [11]. In a flat mesh topology, all the UAVs have the same role, and they work collaboratively. Here, each UAV maintains the topology by interacting with its one- or two-hop members, and the network type is homogenous. Thus, in a flat mesh topology, an optimal multi-hop communication link from a UAV to BS can be established. The flat mesh topology with formation control delivers high tolerance to malfunction and self-healing capability to the entire swarm. Nevertheless, flat mesh topology may incur higher control overhead to maintain the entire swarm topology in a distributed manner. To implement the mmWave-based flat mesh topology, each UAV requires the ability to form the hybrid beamforming to communicate with its multiple neighbor UAVs. Here, the medium access can be shared by adopting the orthogonal frequency division multiple access (OFDMA) [10], time division multiple access (TDMA), frequency division multiple access (FDMA) [12], and space division multiple access (SDMA) [13].

In leader-follower topology-based UAVSNs, a high-capacity leader is responsible to control the topology of the entire swarm by interacting with its followers. The leader can be a high-altitude platform (HAP) such as a high-capacity aircraft or airship. The followers could be small rotary wing UAVs or small fixed-wing UAVs that can execute the mission on a low-altitude platform (LAP). The HAP has higher payload capacity, provides higher endurance and its large platform can support large UPA to form a hybrid beam to the LAP to provide communication coverage. The leader-follower UAVSNs are mostly heterogeneous. The small follower UAVs in LAP can only use the analog beam to communicate with the leader in HAP. The multiple-follower UAVs in LAP can communicate with the leader by sharing the transmission medium using SDMA, non-orthogonal multiple access (NOMA) [7], TDMA, and FDMA. As a major advantage, low-capacity follower UAVs in LAP can offload their computational intensive task to the HAP, to minimize the energy consumption and take faster decision [13]. In addition, HAP has wider coverage to LAP and can obtain a better U2B LoS path [12], [14]. However, because all followers interact with one leader in leader-follower topology, they may encounter communication delay and slower formation convergence.

### 2.4 Challenges in Directional Communication

The mmWave-enabled directional communication in high-mobility UAVSNs encounters huge challenges to perform neighbor discovery, transmission scheduling in the medium access control (MAC) layer, and appropriate relay selection in the routing layer. Because the mmWave beam has limited coverage due to small beamwidth, it is very challenging to discover the neighbor in real time as neighbor UAVs have high mobility. The mmWave quasi-omni-directional NDT has limited coverage owing to small antenna gain. Performing the exhaustive beam scanning in 3D space to perform the neighbor discovery brings delay and higher overhead. Delay in the neighbor discovery process causes source UAVs to deal with the inaccurate position of neighbor UAVs, which may cause severe beam misalignment problems. During transmission scheduling, the directional communication faces deafness (i.e., UAV-B is deaf with respect to UAV-A as UAV-B pointed its beam toward UAV-C in Fig. 1), exposed terminal, and hidden terminal problem in the MAC layer. It happens mainly due to the unheard request to send/clear to send, beam blockage, and asymmetric antenna gain in directional communication, which may cause a collision in data transmission.

## 3 NEIGHBOR DISCOVERY AND BEAM ALIGHMENT TECHNIQUES IN mmWAVE-ENABLED UAVSNS

In this section, the various NDTs and BATs in UAVSNs are reviewed and compared with each other by considering their respective key features, advantages, and limitations.

### 3.1 Neighbor Discovery Techniques

By utilizing NDT, each UAV maintains its one- or two-hop neighbor lists. In a UAVSN, real-time neighbor discovery is required to perform collaborative motion planning, control the swarm topology, avoid the inter-UAV collision, do task allocation, and relay the sensed data to the BS as given in Fig. 1. For beamforming and beam alignment in highly dynamic UAVSNs, it is also very necessary to maintain an updated neighbor list with their precise mobility state (i.e., 3D position, relative velocity, and flying direction) to maximize the link SINR and ensure reliable connectivity. Here, controlling the SI adaptively according to the degree of mobility is required to meet the trade-off between the control overhead and accuracy in successful neighbor discovery [15].

We classified the existing NDT in mmWave-enabled UAVSNs as probabilistic-based, deterministic-based, hybrid-based, dual frequency band-based, and machine learning (ML)-based. In probabilistic-based NDT, each UAV randomly steers its beams to discover the neighbors [16]. In deterministic-based NDT, each UAV steers its beams in a predefined sequence to perform the neighbor discovery [17]. Thus, even though the deterministic approach ensures successful neighbor discovery, it gives a higher average delay compared to the probabilistic approach. In contrast, the probabilistic beam scanning method provides a lower average delay to discover a neighbor. However, in a highly mobile condition, sometimes the probabilistic approach cannot ensure successful neighbor discovery. Considering the trade-off between the lower average delay and higher successful neighbor detection in deterministic and probabilistic-based NDT, the authors in [18] applied the hybrid-based NDT. In hybrid-based NDT, the neighbor discovery process includes deterministic predefined sequences with a random component.

Considering the high mobility of UAVs, the above-mentioned exhaustive searching-based directional neighbor discovery by steering the beam produces a higher delay and low successful neighbor discovery rate. To overcome this problem, dual-frequency band-based NDT is attracted by the researchers [19]. Here, each UAV utilizes low-frequency omni-directional communication for neighbor discovery, controlling the motion, and the medium access. Simultaneously, each UAV utilizes the high frequency mmWave directional communication for data transmission. The low-frequency sub-6 GHz band omni-directional communication has a higher reliable coverage radius as given in Fig. 1, and it can easily solve the challenges in directional communication as discussed in Section 2.4. This technique of separating the control in low frequency and data communication in a high frequency delivers lower delay and higher neighbor discovery success rate. Nevertheless, the interference coming from the external source in the low-frequency sub-6 GHz must be considered as the interference-limited control channel is essential for autonomous UAVSNs.

In ML-based NDT, each UAV acts as a reinforcement learning agent and interacts with the environment to perform the successful neighbor discovery by adopting the Markov decision process. In [20], a Q-learning-based directional NDT was designed by considering the beam steering direction as state, the transmission and reception strategy as action, and the successful neighbor detection as a reward. The ML-based NDT delivers higher successful neighbor discovery and lower average delay. However, it requires training samples to make an intelligent decision. According to the above discussion, the comparative summary of different NDTs are given in Table 1.

### 3.2 Beam Alignment Techniques

After the successful neighbor discovery, UAVs perform the data transmission by aligning the beam toward an appropriate target relay UAV to reach the nearest BS. Owing to the dynamic time-varying topology, frequent NDT produces higher control overhead. In addition, the frequent NDT and target localization process in directional transmission reduces the opportunity for data transmission [4]. After establishing the directional communication between a UAV pair, a small beam misalignment owing to the changes in relative mobility and wind disturbance reduces the link SINR. To minimize the control overhead, maximize the link SINR, and increase the link stability in dynamic UAVSNs, UAVs require to perform continuous beam tracking by predicting the mobility of the selected neighbor relay UAV. This mobility prediction process and performing the beam-steering according to relative mobility to avoid the beam misalignment problem is known as BAT. We classify the BAT algorithm into two categories of Bayesian statistics-based and ML-based techniques.

The source UAV performs the fast beam alignment by predicting the current position of the relay UAV according to its previous position, velocity, acceleration, AoA/AOD of signals, and channel gain. Bayesian statics-based BAT can perform such recursive state estimation by using the Kalman filter and the particle filter. The Kalman filter-based BAT can perform the recursive state estimation in a linear Gaussian environment. The variations of the Kalman filter such as extended Kalman filter (EKF) and unscented Kalman filter (UKF) can perform beam tracking even in non-linear environments [22]. While performing the beam tracking, the previous state information can be acquired by using the onboard sensor (i.e., inertial measurement unit, global positioning system, light detection and ranging, and camera). Additionally, the current position can be predicted according to the previous position, velocity, and acceleration by using the collaborative motion model of UAVSNs [2]. However, in high mobility scenario, localization error must exist between the predicted and actual position.

To obtain a more precise localization and efficient beam tracking, the Kalman filter produces a weighted average between observation and prediction, by adjusting the Kalman gain. However, the Kalman filter can only deal with a linear system. In [4], the Kalman filter-based event-triggered control strategy was applied to maintain the beam alignment to maximize the link SINR and minimize the control overhead. To solve the tracking problem in a non-linear system, EKF introduces the concept of first-order linearization. However, the linearization process in the EKF may not be exactly Gaussian, which brings error in prediction. The UKF further improves the accuracy in the linearization process of the non-linear system to achieve a better mobility prediction [22]. The Kalman filter-based BAT requires a specific system model, which might be non-linear, complex, and less adaptive to the dynamic environment.

The main advantages of ML-based BATs are the adaptive learning capability to adjust to the dynamic time-varying topology without any specific model, and reduce the complexity. The ML-based BAT can be classified further as supervised learning and reinforcement learning techniques. The supervised learning-based BAT includes the long short-term memory (LSTM) and dataset-based ML algorithms. The recurrent neural network such as LSTM can predict the neighbor UAV location using the historical trajectory and channel state information [23]. In [7], the Gaussian process deep neural network algorithm was applied to predict the angular domain information of follower UAVs using a historical dataset to perform the NOMA grouping of follower UAVs, beam tracking, and adaptive beamwidth control. This is for providing higher spectral efficiency and link reliability in leader-follower UAVSNs. However, the major limitation of the supervised learning is that it considers immediate rewards and requires a labeled dataset or historical sequence of trajectory information of UAVs. In contrast, the reinforcement learning considers long-term cumulative rewards, and the agent learns adaptively by interacting with a dynamic environment by adopting the Markov tuple (state, action, and reward) without any prior dataset. In [11], the Q-learning-based BAT for leader-follower UAVSN topology was applied to maximize the link SINR. According to our above discussion, the summary of advantages and limitations of different BATs are given in Table 2.

Table 1: Comparison of different NDTs in mmWave-enabled UAVSNs

| Ref. | NDT type | Key features | Advantages | Limitations |
|---|---|---|---|---|
| [16] | Probabilistic-based | Steers the beam in random direction | Lower neighbor discovery rate | Lower average delay |
| [17] | Deterministic-based | Steers the beam in a predefined sequence | Higher successful neighbor detection rate | Higher average delay |
| [18] | Hybrid-based | Steers the beam in a predefined sequence with a random component. | Considers the trade-off between average delay and successful neighbor discovery | May not be suitable for highly mobile 3D environment |
| [5], [19], and [21] | Dual frequency band-based | Performs (i) low-frequency omni-directional communication to detect neighbor and control the channel access and (ii) high-frequency directional communication only for data transmission. | • Provides lower delay and control overhead<br>• Overcomes the challenges in the MAC and routing layers in directional | External interference in low-frequency channel can affect the performance of |
|  |  |  | communication (i.e., deafness and hidden terminal problems) | the controller. |
| [20] | Machine learning (ML)-based | Each UAV agent interacts with the environment (UAVSN topology) and learns from the experience by considering both the beam steering direction as state and the successful neighbor detection as reward. | Lower average delay and higher successful neighbor discovery rate | Requires higher training cost |

Table 2: Comparison of different BATs in mmWave-enabled UAVSNs

| BAT type | BAT categories | Advantages | Limitations |
|---|---|---|---|
| Bayesian statics-based | Kalman filter [4] | Simple and less complexity | Supports only linear motion model |
|  | Extended Kalman filter (EKF) [24] | Less complexity compared to UKF | Includes error in linear approximation of non-linear systems |
|  | Unscented Kalman filter (UKF) [22] | More precise mobility prediction due to higher accuracy in linearization | Only works for Gaussian environment |
| Machine learning (ML)-based | Supervised learning (LSTM [23], dataset-based [7]) | Higher accuracy in mobility prediction | Requires labeled dataset and long training time, and considers short-term reward |
|  | Reinforcement learning (Q-learning [11]) | Adaptable to dynamic environment and considers long-term cumulative reward | Requires training samples to make better mobility prediction |

## 4 OPEN ISSUES AND CHALLENGES

In this section, according to our review in Section 3, we discuss important open issues and research challenges.

### 4.1 Dual Frequency Operation for Control and Communication

Owing to the higher communication radius, the sub-6 GHz low-frequency omni-directional communication is very useful for performing neighbor discovery, controlling the mobility, performing beam alignment, and scheduling the MAC layer transmission. It helps to avoid the delay in neighbor discovery, deafness, and hidden terminal problem in directional communication. Then, high-frequency mmWave directional communication only for data transmission can ensure a high data rate. The dual-antenna operation in dynamic mmWave-enabled UAVSNs requires further study to separate the control panel in low frequency and the data transmission in high frequency [5], [21].

### 4.2 Cross-Layer Design

In mmWave directional communication, the mobility control, mobility prediction, beamforming, alignment, adaptive beamwidth control in the physical layer, and MAC-layer allocation of resources (i.e., transmission power, frequency, or timeslot) are highly coupled to maximize the link SINR. Thus, the cross-layer routing protocol design in mmWave-enabled UAVSN is highly required, which is rarely studied in the existing literature.

### 4.3 Hierarchical Aerial Edge Computing

Offloading the computation-intensive tasks from low-capacity follower UAVs in LAP to high-capacity leader UAVs in HAP helps to minimize the energy consumption of LAP. In addition, the HAP provides shorter communication delay as the LAP to HAP communication path offers a pure LoS link. Thus, it is needed to develop the hierarchical aerial edge computing solution in mmWave-enabled UAVSNs [7], [13].

### 4.4 Joint Optimizations of Control and Communication

In mmWave-based directional communication, a stable communication link is required to perform the beamforming and to maintain the beam alignment. Owing to high mobility, wind disturbance, and limited beam coverage, a slight beam misalignment creates a link outage in directional communication. The flocking control strategy inspired by boids flocking, artificial potential field, and consensus-based graph theory can help UAVSNs to maintain the stable formation in a dynamic environment with minimal control overhead [2]. Thus, the joint study of the stable formation using control theory and mmWave beamforming is necessary in the future.

### 4.5 Mobility Prediction Strategy in Dynamic UAVSNs

In the high-mobility UAVSNs, the mmWave-based directional communication requires the beam alignment to maintain the link stability with desired SINR. To implement the adaptive beam alignment between transceivers, the mobility prediction strategy is required. The event trigger control strategy with Kalman filter-based and LSTM-based mobility prediction (using historical UAV trajectory information, AoA/AoD of signal, and channel state information) facilities better accuracy in beam alignment. This is not only for maximizing link SINR and data transmission opportunity but also for minimizing beam training overhead. Thus, further study should be devoted to designing intelligent algorithms jointly considering the mobility prediction, sensing (controlling the adaptive beamwidth and SI), and communication [4].

### 4.6 IRS-Assisted mmWave Communication in UAVSNs

The mmWave signals face higher path loss in long distances and no LoS condition. The IRS-assisted passive beamforming can overcome this problem by creating a virtual LoS path to improve the channel gain with lower energy and less hardware complexity. Thus, the on-ground IRS-assisted mmWave communication in UAVSNs can facilitate UAVs to establish a better U2G link, which requires further study.

## 5 CONCLUSIONS

In this paper, we have summarized the basic requirement to implement the mmWave-enabled UAVSNs by addressing the antennas, beamforming, topology formation, and challenges in directional communication. We have surveyed the different NDTs and BATs in mmWave-enabled UAVSNs along with their key features, advantages, and limitations. Finally, we have discussed the key open issues and challenges with potential research direction in this field.

## ACKNOWLEDGMENTS

This work was supported in part by the National Research Foundation of Korea (NRF) grant funded by the Korea government (MIST) (No. 2022R1A2C1009037). All correspondence should be addressed to Sangman Moh (smmoh@chosun.ac.kr).